\documentclass[a4paper,11pt]{article}
\usepackage[section]{placeins}

\usepackage{graphicx}
\usepackage{dcolumn}
\usepackage{bm}
\usepackage{multirow}
\usepackage{array}
\usepackage{tabularx}
\usepackage{xcolor}
\usepackage{lipsum}
\usepackage{enumitem}
\usepackage{setspace}
\usepackage{authblk}
\usepackage{cite}
\usepackage[compact]{titlesec} 
\usepackage{geometry}
\usepackage{amsmath}
\usepackage[mathlines]{lineno}
\title{Longitudinal double spin asymmetry of $\pi^{\pm}$-tagged jet, $\Lambda$, $\overline{\Lambda}$, and $K_S^0$ in polarized $p+p$ collisions at $\sqrt{s}=200$ $\rm{GeV}$ at STAR}
\author{Yi Yu, for the STAR Collaboration \\ Institute of Frontier and Interdisciplinary Science \& Key Laboratory of Particle Physics and Particle Irradiation (Ministry of Education), Shandong University, Qingdao, Shandong 266237, China}

\begin{document}
\maketitle
Understanding the origin of the proton spin is one of the most fundamental and challenging questions in QCD. Much progress has been made since the first surprising result by the EMC experiment in the late 1980s. However, the helicity distributions of strange quarks and anti-quarks inside the proton are still not well constrained by the experimental data.
Measurement of the longitudinal double spin asymmetry $A_{LL}$ of the inclusive jets tagged with a $\pi^{\pm}/\pi^-$ carrying high jet momentum fraction, $z$, in $p+p$ collisions can provide further constraints on the gluon helicity distribution in the proton.
In addition, the $A_{LL}$ of $\Lambda$, $\overline{\Lambda}$ and $K_S^0$ in the longitudinally polarized $p+p$ collisions may shed light on the strange quark and anti-quark helicity distributions.
In this contribution, we report the preliminary results on the $A_{LL}$ measurements of inclusive jets tagged with a high-$z$ $\pi^{\pm}$, and the $\Lambda$, $\overline{\Lambda}$ and $K_S^0$. We utilize the longitudinally polarized $p+p$ collisions at $\sqrt{s}=200$ $\rm{GeV}$ collected by the STAR experiment with an integrated luminosity of about 52 $\rm{pb^{-1}}$.



\section{Introduction}
Inspired by the first surprising results of the proton spin structure from the EMC Collaboration\,\cite{Ashman_1988_MeasurementSpin}, tremendous progress has been made in the past 35 years to improve our knowledge of how the proton spin is made up by the quarks and gluons in Quantum Chromodynamics (QCD). Nevertheless, there still remain open questions that challenge the current understanding of the composition of the proton spin. In polarized $p+p$ collisions, the longitudinal double spin asymmetry $A_{LL}$ is expected to be sensitive to the helicity distributions of partons inside the proton, which is defined as the difference between the cross section with same and opposite beam helicity:
\begin{align}
    A_{LL}\equiv \frac{\sigma^{++}-\sigma^{+-}}{\sigma^{++}+\sigma^{+-}}=\frac{\Delta \sigma}{\sigma}
\end{align}
At leading order, the $\Delta \sigma$ is proportional to $\Delta f_1\Delta f_2\hat{a}_{LL}$, where $\Delta f_1$ and $\Delta f_2$ are the helicity distributions of two colliding partons and $\hat{a}_{LL}$ is the double spin asymmetry at partonic level and can be calculated by the perturbative QCD.

The Relativistic Heavy Ion Collider (RHIC) is the world's first and only polarized $p+p$ collider and is capable of colliding both longitudinally and transversely polarized proton beams at $\sqrt{s}=200$ GeV and 500 GeV. The dominant subprocesses of the hard scattering in such $p+p$ collisions are quark-gluon and gluon-gluon scatterings\,\cite{jetALL_2019}, which make RHIC an ideal facility to study the gluon helicity distribution. Series of $A_{LL}$ measurements\,\cite{ jetALL_2019,jetALL_2015, jetALL_2018, jetALL_2021, jetALL_2022, dijet_2023} for jets and dijets have confirmed a sizable gluon polarization\,\cite{DSSV_2014,nnpdfpol1.1} inside the longitudinally polarized proton. However, the JAM Collaboration recently proposed that $A_{LL}$ of inclusive jets is not sensitive to the sign of the gluon helicity distribution and a negative gluon polarization is also allowed\,\cite{JAM2022}. As the helicity distributions of $u$ quark and $d$ quark are in opposite sign, and they favor $\pi^+$ and $\pi^-$ in the hadronization, respectively, $A_{LL}$ of the $\pi^{\pm}$-tagged jets is expected to be sensitive to the sign of the gluon helicity. In addition, the helicity distributions of the strange quark and anti-quark\,\cite{nnpdfpol1.1, Ethier_2017_FirstSimultaneous} are still poorly constrained by the experiments. Measuring the longitudinal double spin asymmetry of $\Lambda$ hyperons and $K_S^0$ can shed light on the strange quark and anti-quark helicity distributions.

In 2015, STAR concluded its longitudinal polarized data collection at $\sqrt{s}=200$ GeV. This data set corresponds to an integrated luminosity of about 52 $\mathrm{pb}^{-1}$ with an average beam polarization of about 54\%. With this data set, we performed the first measurement of the $A_{LL}$ for the $\pi^{\pm}$-tagged jet, $\Lambda$ hyperons, and $K_S^0$.
\section{Longitudinal Double Spin Asymmetry $A_{LL}$ of $\pi^{\pm}$-tagged Jet}
\subsection{Jet Reconstruction}
Similar to previous publications from STAR\,\cite{jetALL_2019, jetALL_2021, jetALL_2022}, jets were reconstructed with charged-particle tracks measured by the Time Projection Chamber (TPC)\,\cite{TPC} and the energy deposits in the Barrel Electromagnetic Calorimeter (BEMC)\,\cite{BEMC} and the Endcap Electromagnetic Calorimeter (EEMC)\,\cite{EEMC} using the anti-$k_T$ algorithm\,\cite{antikt} with a resolution parameter $R=0.6$. An off-axis cone method adapted from the ALICE experiment\,\cite{offaxis} was used to correct the transverse momentum $p_T$ of the reconstructed jet for the contribution of the underlying events. A Monte Carlo simulation sample was generated to correct the reconstructed jet quantities back to particle level and estimate systematic uncertainties in these measurements. In the simulation, the $p+p$ collision events were generated with PYTHIA6\,\cite{pythia6} and were further processed through a STAR detector-response package based on GEANT3\,\cite{geant3}. The simulated events were then embedded into the zero-bias events recorded during the STAR data-taking runs to simulate the pile-ups and beam backgrounds present in data. Same jet reconstruction procedure was applied to this simulation sample. In addition, particle-level jets were reconstructed and the jet $p_T$ in data was corrected back to particle level.
\subsection{$\pi^{\pm}$ Identification}
In this measurement, charged pions were identified based on their energy loss $dE/dx$ inside the TPC. At STAR, the $dE/dx$ is normalized into $n\sigma$ according to the following formula:
\begin{align}
    n\sigma(\pi) = \frac{1}{\sigma_{\mathrm{exp}}}\mathrm{ln}
    \left(\frac{dE/dx_{\mathrm{obs}}}{dE/dx_{\pi,\mathrm{cal}}}\right),
\end{align}
where the $dE/dx_{\rm{obs}}$ and the $dE/dx_{\pi, \rm{cal}}$ are the measured energy loss of the tracks and the expected energy loss of the $\pi^{\pm}$ based on the Bichsel formalism\,\cite{BICHSEL}. The $\sigma_{\rm{exp}}$ is the energy loss resolution of the TPC\,\cite{XU201028,SHAO2006419}. 
\begin{figure}[htbp]
    \centering
    \includegraphics[width=1\linewidth]{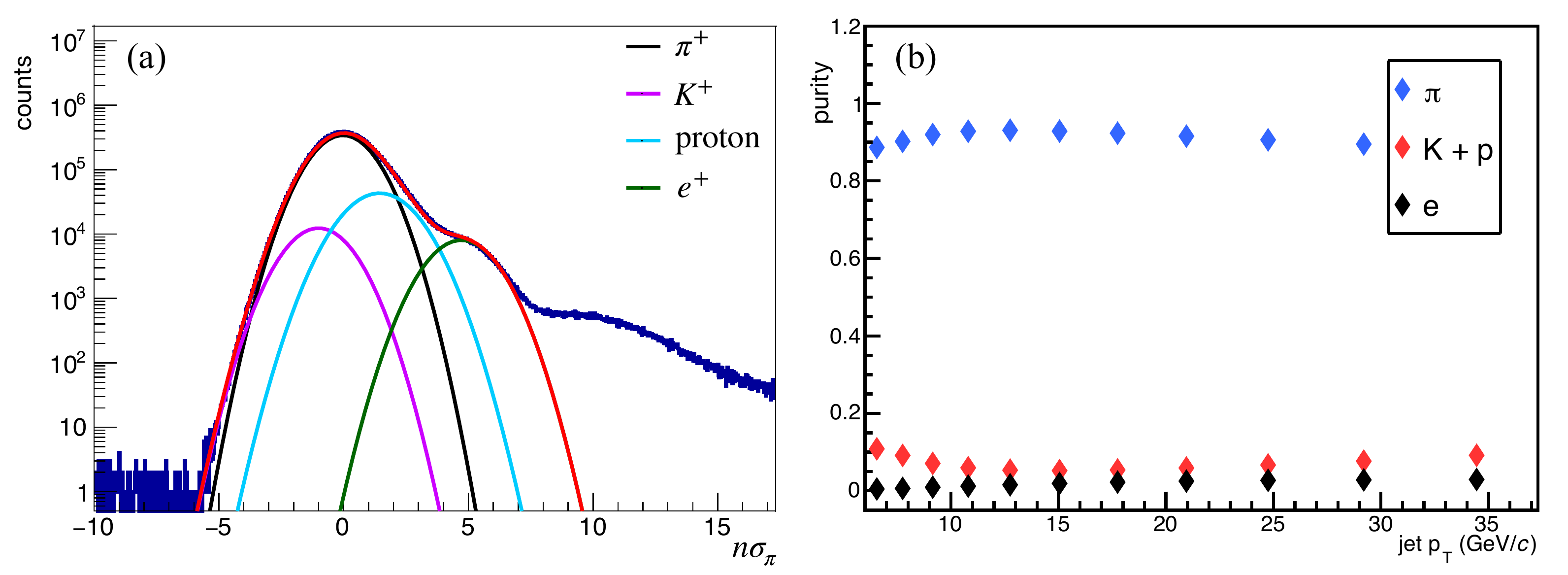}
    \caption{(a) Multi-Gaussian fitting on the $n\sigma(\pi)$ distribution for positive charged tracks with their momentum $1.4<p<1.5$ GeV$/c$. (b) Particle purity as a function of jet transverse momentum $p_T$ in pion-rich region for positive charged tracks carrying jet momentum fraction $z>0.2$.}
    \label{fig:pion_pid}
\end{figure}
In practice, the $n\sigma(\pi)$ distribution was divided into three particle-rich regions, i.e., pion-rich region, kaon+proton-rich region, and electron-rich region. Similar to \,\cite{collins_2015}, particle purity at each particle-rich region was estimated with a multi-Gaussian fitting to the $n\sigma(\pi)$ distribution of charged tracks. Figure\,\ref{fig:pion_pid}(a) shows an example of the multi-Gaussian fitting of the $n\sigma(\pi)$ distribution for positive charged tracks with their momentum $1.4<p<1.5$ GeV$/c$. The estimated purity, for example, as a function of jet $p_T$ in pion-rich region is presented in Fig.\,\ref{fig:pion_pid}(b). To enhance the fraction of jets from the fragmentation of $u$ quark and $d$ quark, jet momentum fraction $z\equiv \overrightarrow{p}_{\pi}\cdot \overrightarrow{p}_{jet}/|\overrightarrow{p}_{jet}|^2$ carried by the $\pi^{\pm}$ was required to be larger than 0.2.
\subsection{$A_{LL}$ Extraction \& Result}
Experimentally, the longitudinal double asymmetry $A_{LL}$ can be calculated with the following formula:
\begin{align}
    A_{LL}=\frac{1}{P_Y P_B}\frac{N^{++} - \mathcal{R}N^{+-}}{N^{++} + \mathcal{R}N^{+-}},
    \label{eq:ALLextraction}
\end{align}
where $P_Y$ and $P_B$ are the measured polarization of two colliding beams. $N^{++}$ and $N^{+-}$ are the $\pi^{\pm}$-tagged jet yields from beam bunches with same and opposite beam helicity configurations, respectively. $\mathcal{R}$ is the relative luminosity measured by the Vertex Position Detectors (VPD)\,\cite{VPD} and the Zero Degree Calorimeters (ZDC)\,\cite{ZDC}.
The raw asymmetry, $A_{LL}^{\rm{raw}}$, calculated from each particle rich region is expected to be a linear combination of the $A_{LL}$ of each particles. The $\pi^{\pm}$-tagged $A_{LL}$ can be obtained by solving the following linear equations:
\begin{align}
    \begin{bmatrix}
f_{\pi_{\rm{rich}}}^{\pi} & f_{\pi_{\rm{rich}}}^{K+p} & f_{\pi_{\rm{rich}}}^{e} \\[1mm]
f_{K+p_{\rm{rich}}}^{\pi} & f_{K+p_{\rm{rich}}}^{K+p} & f_{K+p_{\rm{rich}}}^{e} \\[1mm]
f_{e_{\rm{rich}}}^{\pi} & f_{e_{\rm{rich}}}^{K+p} & f_{e_{\rm{rich}}}^{e}
\end{bmatrix}
\begin{bmatrix}
A_{LL}^{\pi} \\[1mm] A_{LL}^{K+p} \\[1mm] A_{LL}^{e}
\end{bmatrix}
=
\begin{bmatrix}
A_{LL}^{{\rm{raw}}, \pi_{\rm{rich}}} \\[1mm] A_{LL}^{{\rm{raw}}, K+p_{\rm{rich}}} \\[1mm] A_{LL}^{{\rm{raw}}, e_{\rm{rich}}}
\end{bmatrix},
\end{align}
where, for example, $f_{\pi_\mathrm{\rm{rich}}}^{\pi}$ is the $\pi^{\pm}$ purity at $\pi^{\pm}$ rich region.

\begin{figure}[htbp]
    \centering
    \includegraphics[width=1\linewidth]{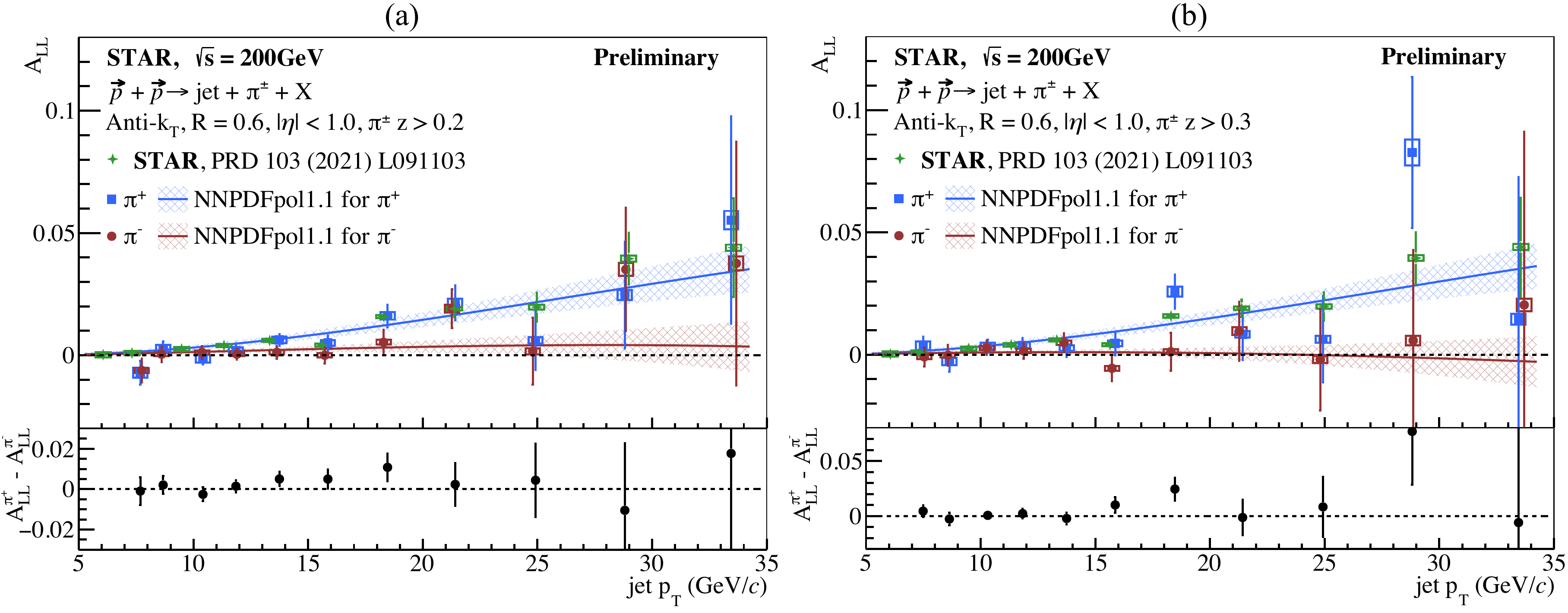}
    \caption{Panel (a) and (b): preliminary results of the longitudinal double spin asymmetry $A_{LL}$ of the $\pi^{\pm}$-tagged jet as a function of jet $p_T$ in polarized $p+p$ collisions at $\sqrt{s}=200$ GeV with jet momentum fraction $z>0.2$ and $z>0.3$, respectively. The bars represent the statistical uncertainties while the systematic uncertainties are shown in boxes. Predictions calculated with the helicity distributions\,\cite{nnpdfpol1.1} based on PYTHIA6\,\cite{pythia6} are compared to data.}
    \label{fig:pion_ALL}
\end{figure}

The preliminary results of the $\pi^{\pm}$-tagged jet $A_{LL}$ with jet momentum fraction $z>0.2$ and $z>0.3$ are presented in Fig.\,\ref{fig:pion_ALL} (a) and Fig.\,\ref{fig:pion_ALL} (b), respectively. Predictions based on PYTHIA6\,\cite{pythia6} and the helicity distributions from NNPDF Collaboration\,\cite{nnpdfpol1.1} are compared with the measurements. The measurements are consistent with the predictions of $A_{LL}^{\pi^+}>A_{LL}^{\pi^-}$ with positive gluon helicity.

\section{Longitudinal Double Spin Asymmetry of $\Lambda$, $\overline{\Lambda}$ and $K_S^0$}
\subsection{$\Lambda$, $\overline{\Lambda}$ and $K_S^0$ Reconstruction}
\begin{figure}[htbp]
\centering
\includegraphics[width=1\linewidth]{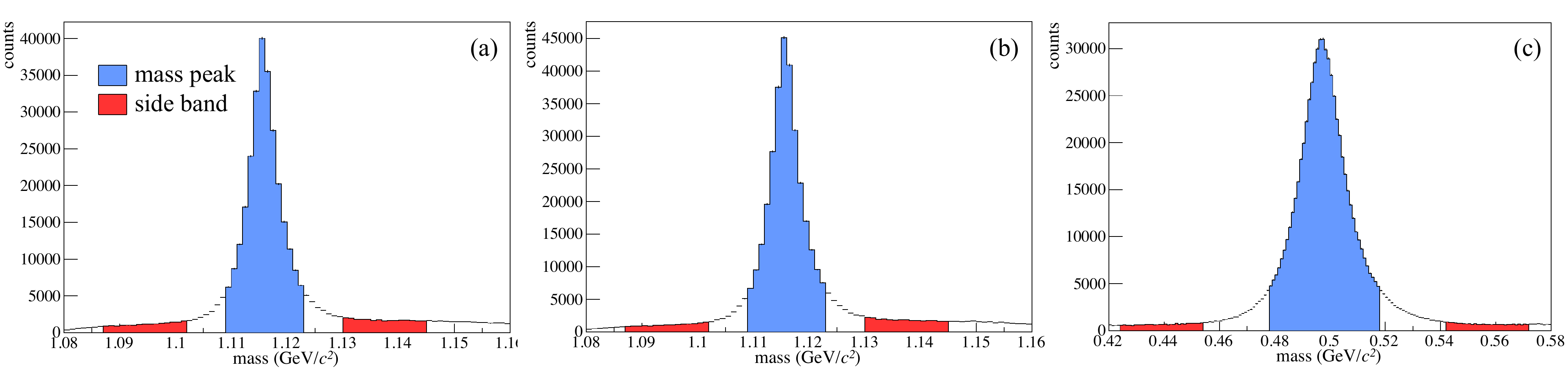}
\caption{Panel (a), (b) and (c) show the invariant mass distributions of the reconstructed $\Lambda$, $\overline{\Lambda}$ and $K_S^0$ candidates at $3<p_T<4$ GeV$/c$, respectively. 
The yields under the mass peak (blue-filled area) are used for the $A_{LL}^{raw}$ calculation 
and the yields under the side-band region (red-filled area) are used for estimating the background fraction under the mass peak.}
\label{fig:mass}
\end{figure}

The $\Lambda(\overline{\Lambda})$ hyperons and $K_S^0$ were reconstructed via their decay channels $\Lambda \to p + \pi^{-}$ $(\overline{\Lambda}\to \overline{p} + \pi^+)$, and $K_S^0\to \pi^+ + \pi^-$, respectively. Daughter tracks were identified based on their energy loss $dE/dx$ inside the TPC. Two daughter candidates were paired firstly and a set of selection criteria was applied based on the decay topology of the $\Lambda$ hyperons and $K_S^0$. Figure\,\ref{fig:mass} shows the invariant mass distribution of the reconstructed $\Lambda$, $\overline{\Lambda}$, and $K_S^0$ candidates at $3<p_T<4$ $\mathrm{GeV}/c$. The residual backgrounds under their mass peak are mainly from the random combination of two daughter candidates and were estimated with the side-band method.


The reconstructed $\Lambda$, $\overline{\Lambda}$ and $K_S^0$ candidates were used as inputs in the jet reconstruction using the same method as in Ref.\,\cite{spintransfer_2023}. In-jet $\Lambda$ hyperons and $K_S^0$ were used for further analysis to make sure $\Lambda$ hyperons and $K_S^0$ originate from a hard partonic scattering.

\begin{figure*}[htbp]
    \centering
    \includegraphics[scale=0.6]{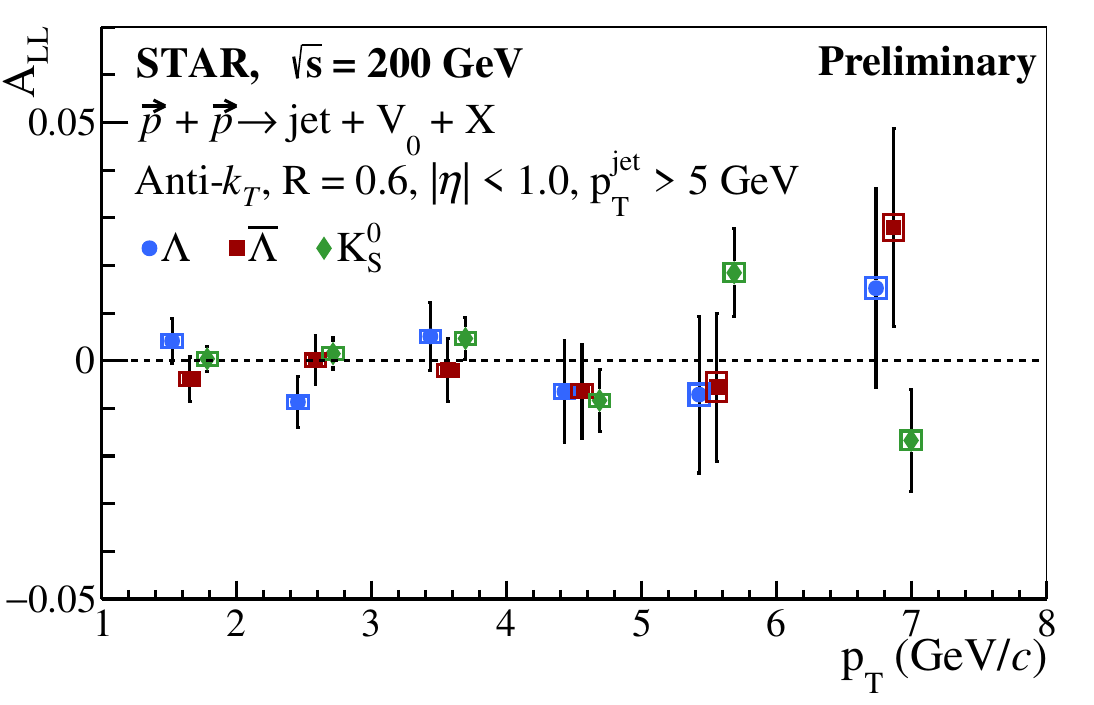}
    \caption{Preliminary results of longitudinal double spin asymmetry $A_{LL}$ as a function of particle $p_T$ for $\Lambda$, $\overline{\Lambda}$, and $K_S^0$ in polarized $p+p$ collisions at $\sqrt{s}=200$ GeV. The results of $\overline{\Lambda}$ and $K_S^0$ have been slightly offset horizontally for clarity.}
    \label{fig:LAK_ALL_pt}
\end{figure*}
\begin{figure*}[htbp]
    \centering
    \includegraphics[scale=0.6]{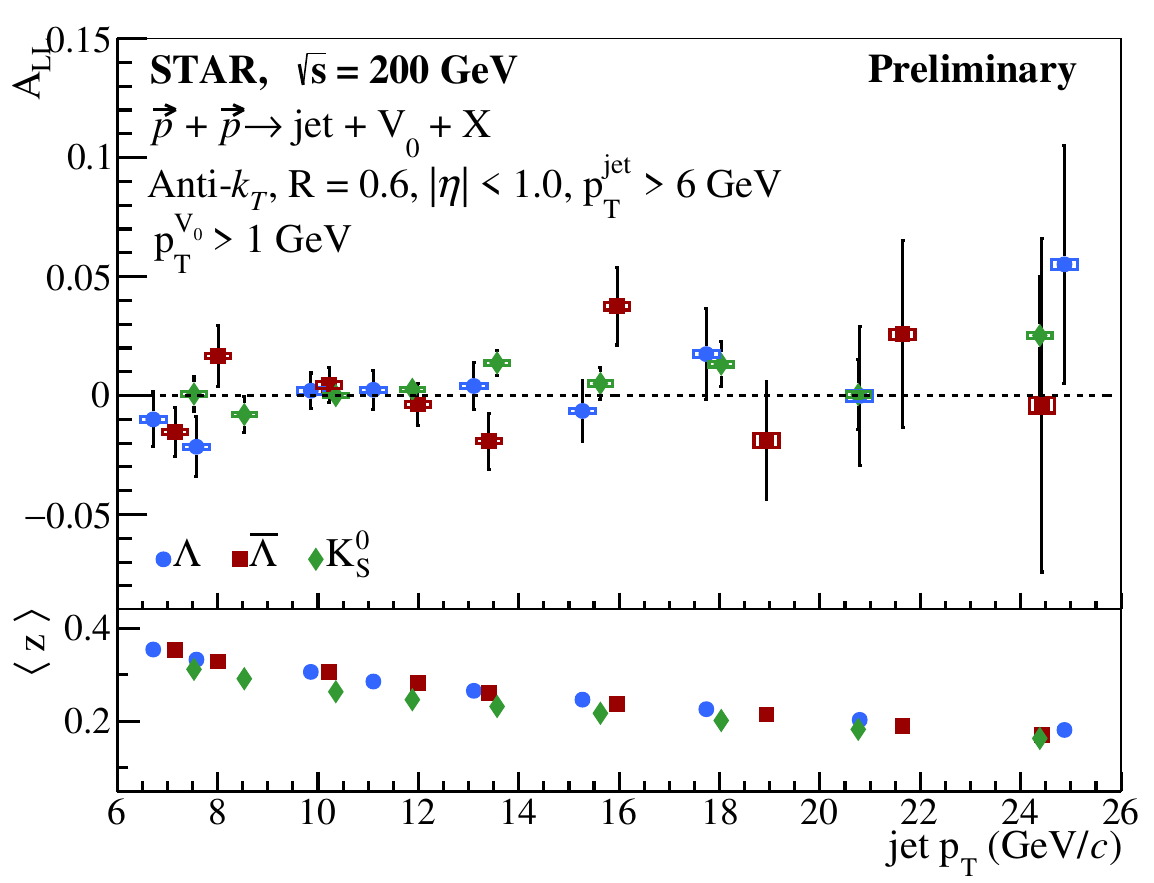}
    \caption{Top panel: preliminary results of longitudinal double spin asymmetry $A_{LL}$ as a function of jet $p_T$ for $\Lambda$, $\overline{\Lambda}$, and $K_S^0$ in polarized $p+p$ collisions at $\sqrt{s}=200$ GeV. Bottom panel: the averaged particle-level jet momentum fraction $z$ carried by $\Lambda$ hyperons and $K_S^0$.}
    \label{fig:LAK_ALL_jetpt}
\end{figure*}
\subsection{$A_{LL}$ Extraction \& Result}
Similarly, the longitudinal double spin asymmetry of the $\Lambda$ hyperons and the $K_S^0$ can be measured with the Eq.\,(\ref{eq:ALLextraction}).
The $A_{LL}^{\rm{raw}}$ from mass peak and the $A_{LL}^{\rm{bkg}}$ from the side-band region were extracted separately with the Eq.\,(\ref{eq:ALLextraction}). The influence of the residual backgrounds was corrected with the following formula:
\begin{align}
    A_{LL}=\frac{A_{LL}^{\rm{raw}}-rA_{LL}^{\rm{bkg}}}{1 - r},
    \label{eq:LAK_ALL2}
\end{align}
where $r$ is the residual background fraction under mass peak and is found to be below 10\%.

The preliminary results of the longitudinal double spin asymmetry of the in-jet $\Lambda$ hyperons and $K_S^0$ as a function of particle $p_T$ are presented in Fig.\,\ref{fig:LAK_ALL_pt}. The results are consistent with zero within uncertainties. Figure \,\ref{fig:LAK_ALL_jetpt} shows the $A_{LL}$ as a function of the jet $p_T$. The jet momentum fraction $z$ shown on the bottom panel was corrected back to particle level with the same Monte Carlo simulation samples used in $\pi^{\pm}$-tagged jet $A_{LL}$ measurements. The jet sample used in Fig.\,\ref{fig:LAK_ALL_jetpt} has large overlap with the published inclusive jet $A_{LL}$ measurement\,\cite{jetALL_2021}, but this result is more sensitive to the strange quark and anti-quark helicity distributions as $\Lambda$ hyperons and $K_S^0$ are a part of jet. Another related analysis is the longitudinal spin transfer of the $\Lambda$ hyperons, which is also sensitive to the strange quark and anti-quark helicity distributions\,\cite{spintransfer_2023}. Small $A_{LL}$ results might indicate small strange quark and anti-quark helicity distributions inside the proton.

\section{Summary}
We reported new preliminary results of the longitudinal double spin asymmetry $A_{LL}$ of the $\pi^{\pm}$-tagged jet, $\Lambda$ hyperons, and $K_S^0$ in polarized $p+p$ collisions at $\sqrt{s} = 200$ GeV at STAR. The $\pi^{\pm}$-tagged jet $A_{LL}$ can provide sensitivity to the sign of the gluon helicity distribution. The results indicate $A_{LL}^{\pi^+}>A_{LL}^{\pi^-}$, which favor positive gluon helicity. The first measurement of $A_{LL}$ for the in-jet $\Lambda$ hyperons and $K_S^0$ is consistent with zero within uncertainties, which indicates small helicity distribution of the strange quark and anti-quark inside the proton.
\section*{Acknowledgements}
The author is supported partially by the National Natural Science Foundation of China under No. 12075140.
\bibliography{ref.bib}

\end{document}